\documentclass[fleqn,usenatbib]{mnras}

\usepackage{newtxtext,newtxmath}

\usepackage[T1]{fontenc}

\DeclareRobustCommand{\VAN}[3]{#2}
\let\VANthebibliography\thebibliography
\def\thebibliography{\DeclareRobustCommand{\VAN}[3]{##3}\VANthebibliography}

\usepackage{graphicx}	
\usepackage{amsmath}	

\def\Cell{$C_{\ell}$ }
\def\3x2pt{3$\times$2pt}
\newcommand{\tens}[1]{\mathbfss{#1}}
\newcommand{\bm}[1]{\bmath{#1}}

\title[On cosmic shear without lensing bias]{Avoiding lensing bias in cosmic shear analysis}

\author[C. A. J. Duncan \& M. L. Brown]{
Christopher A. J. Duncan,$^{1}$\thanks{E-mail: christopher.duncan@manchester.ac.uk} \&
Michael L. Brown$^{1}$\thanks{E-mail: m.l.brown@manchester.ac.uk}
\\
$^{1}$Jodrell Bank Centre for Astrophysics, Department of Physics and
Astronomy, University of Manchester, Oxford Road, Manchester
M13 9PL, United Kingdom\\
}

\date{Accepted XXX. Received YYY; in original form ZZZ}

\pubyear{\the\year{}}

\begin{document}
\label{firstpage}
\pagerange{\pageref{firstpage}--\pageref{lastpage}}
\maketitle

\begin{abstract}
We show, using the pseudo-$C_\ell$  technique, how to estimate cosmic shear and galaxy-galaxy lensing power spectra that are insensitive to the effects of multiple sources of lensing bias including source-lens clustering, magnification bias and obscuration effects. All of these effects are of significant concern for ongoing and near-future Stage-IV cosmic shear surveys. Their common attribute is that they all introduce a cosmological dependence into the selection of the galaxy shear sample. Here, we show how a simple adaptation of the pseudo-$C_\ell$ method can help to suppress these biases to negligible levels in a model-independent way. Our approach is based on making pixelised maps of the shear field and then using a uniform weighting of those shear maps when extracting power spectra. To produce unbiased measurements, the weighting scheme must be independent of the cosmological signal, which makes the commonly-used inverse-variance weighting scheme unsuitable for cosmic shear measurements. We demonstrate this explicitly. A frequently-cited motivation for using inverse-variance weights is to minimize the errors on the resultant power spectra. We find that, for a Stage-IV-like survey configuration, this motivation is not compelling: the precision of power spectra recovered from uniform-weighted maps is only very slightly degraded compared to those recovered from an inverse-variance analysis, and we predict no degradation in cosmological parameter constraints. We suggest that other 2-point statistics, such as real-space correlation functions, can be rendered equally robust to these lensing biases by applying those estimators to pixelised shear maps using a uniform weighting scheme. 
\end{abstract}




\section{Introduction}
\label{sec:intro}
Weak gravitational lensing, utilising the observed positions and shapes of a large sample of galaxies, has become one of the cornerstones of modern cosmological inference, culminating in the recent analyses of Stage-III surveys such as the Hyper Suprime Cam Survey \citep{Hikage2019HSC, Hamana20202PCF, dalal2023}, the Dark Energy Survey \citep{Abbott2022DES, Amon2022DES, Secco2022DES}, and the Kilo-Degree Survey \citep{Heymans2021KiDS1000, Asgari2021KiDS1000}. Forthcoming data from current Stage-IV surveys such as those from the \emph{Euclid} satellite \citep{Euclid2024Overview} and the Vera Rubin telescope \citep{Ivezi2019LSST} will provide a large step forward in both the quantity and quality of available weak lensing data, further improving the statistical power of the analysis of their data but, as a result, requiring ever-tighter control of systematic effects throughout the pipeline.

One category of systematics is those that introduce a correlation between the observed local number density of galaxies and the cosmological signal. This includes the so-called source-lens clustering (SLC) effect \citep{Bernardeau1999SLC, Hamana2002SLC, Schmidt2009SLC, Yu2015, Deshpande2024, Linke2024SLC}, arising due to the fact the shear field can only be measured at the positions at which galaxies are observed. Since the distribution of observed galaxies on the sky is, in part, determined by the overall matter density field, this means that line-of-sight positions associated with large mass over-densities (and hence, large shears) are sampled more often than those associated with under-densities (small shears), leading to a selection-effect type bias in the measured shear field. Secondly, lensing magnification \citep{turner1984, Schmidt2009SLC} results in additional correlations between the shear field and the source number density fluctuations, the exact details of which are dependent on the galaxy luminosity function and the galaxy sample selection function. Thirdly, source obscuration effects will be more significant along lines-of-sight that traverse high-mass regions, again introducing additional correlations between the source number density and shear fields \citep{hartlap2009}. Hereafter, we refer to all three of these effects simply as ``lensing bias''. 

The combined effect of these phenomena on weak lensing statistics is difficult to model analytically, and subject to significant uncertainty. Moreover, these effects are of significant concern for current and future Stage-IV surveys \citep{Deshpande2020, duncan2022, Deshpande2024}. For example, after considering the potential impact of 24 possible systematic effects for weak lensing shear analyses with \emph{Euclid}, \cite{Deshpande2024} identified these three effects -- SLC, magnification bias of the source sample, and source obscuration -- as the most concerning. 

Cosmological analyses of large photometric surveys rely primarily on the estimation and interpretation of combinations of 2-point correlations, between observed galaxy positions (galaxy clustering), galaxy shear or shape (cosmic shear), and cross-correlations between position and shear (galaxy-galaxy lensing). Headline cosmological inference results from state-of-the-art surveys are now routinely presented using so-called ``3$\times$2pt'' analyses combining all three. 

These 2-point correlations can be constructed either in real space, using correlations as a function of physical separation, or in harmonic space under a harmonic transformation. Increasingly, survey collaborations are measuring two-point statistics in both real and harmonic space \citep[e.g.][]{Amon2022DES, doux2022, Asgari2021KiDS1000, Loureiro2022, Hikage2019HSC}, with agreement between different statistics providing an important verification of the measured signal and inferred cosmological results \citep[e.g.][]{doux2021, Asgari2021KiDS1000, dalal2023}.

In this paper, we demonstrate how one can render cosmic shear and 3$\times$2pt analyses insensitive to lensing bias through a simple adaptation of the process by which the two-point statistics are extracted. We demonstrate our approach using the popular pseudo-$C_\ell$ power spectrum estimator \citep{peebles1973}, as implemented via the ``MASTER'' algorithm, first developed in the context of Cosmic Microwave Background (CMB) analyses \citep{hivon2002, kogut2003, Brown2005PsClCMB}. This approach was later adapted and applied to weak lensing analyses \citep[e.g.][]{hikage2011, demetroullas2016, asgari2018, Hikage2019HSC, doux2022}. A popular implementation of the MASTER algorithm is the publicly available \textsc{NaMaster} software package \citep{AlonsoNaMaster, AlonsoNamasterPaper}.

Traditionally, pseudo-$C_\ell$ estimators are applied to pixelised maps of spin-0 and spin-2 fields on the sky, such as the CMB total intensity and polarisation fields, or the galaxy clustering and weak lensing shear fields. Pixelised maps are a natural choice for CMB analyses due to the continuous, and band-limited, nature of the CMB fields on the sky. More recently, several authors have proposed adaptations of the pseudo-$C_\ell$ approach such that it can be applied directly to discretely sampled fields \citep{baleato2024, Wolz24, Tessore24}. These adaptations, which are aimed primarily at late-time large-scale structure analyses, are motivated by the fact that the galaxy over-density and weak lensing fields are discretely sampled at the observed galaxy positions.  

An important analysis choice when measuring pseudo-$C_\ell$ power spectra from pixelised maps is the weighting scheme one applies to the maps. Inverse-variance weighting is often recommended \citep[e.g.][]{Nicola21, Tessore24, Alonso2024} and, to date, pseudo-$C_\ell$ weak lensing analyses have typically adopted this approach. For example, \cite{demetroullas2016, Hikage2019HSC, doux2022}  and \cite{dalal2023} all used inverse-variance (or equivalent) weighting schemes. An exception is \cite{Loureiro2022}, who used a uniform weighting scheme, and highlighted potential issues with the use of a weighting scheme based on the per-galaxy shear weights imprinting unaccounted-for information from the clustering of galaxies on the pseudo-$C_\ell$ analysis. A further potential issue with the use of shear weights in a pixel weighting scheme is that the measurement uncertainties for each source in a weak lensing catalogue are themselves estimates, and may be subject to systematic bias which may propagate through the pseudo-\Cell estimator, for example due to correlations between the estimated shear on a source image and systematics as part of the image analysis \citep[e.g.][]{Giblin2021}. 

The generally accepted rule-of-thumb for CMB analyses is that an inverse-variance weight is optimal in the noise-dominated r\'egime while a uniform weight is preferred in the case of signal-dominated maps \cite[e.g.][]{2004MNRAS.349..603E}.\footnote{For weak lensing and \3x2pt analyses, given the wide range of angular scales and redshifts probed by Stage-IV surveys, it is likely that a correspondingly wide range of signal-to-noise r\'egimes will be present in the data, and therefore that it will not be possible to identify a single ``optimal'' weighting scheme.} Note that the ``optimal weight'' in this context means the weighting scheme that produces power spectrum estimates with the smallest error bars. For CMB observations, the variance in map pixel values depends only on instrumental factors such as the integration time and instrumental systematics. However, for weak lensing, the variance in map pixel values will depend on the number of galaxies observed in each pixel, and therefore on the cosmological signal, in addition to observational factors. This dependence of the map pixel variance on the cosmological signal means that an inverse-variance weighting scheme will result in biased estimates of the cosmic shear and galaxy-galaxy lensing power spectra.

In this paper, we demonstrate that the pseudo-$C_\ell$ pixel weighting scheme is intimately connected to lensing bias and that the lensing bias can be effectively removed, at no cost,  through an appropriate choice of weighting scheme. We do this by comparing power spectrum analyses of simulations of the \3x2pt fields using both inverse-variance and uniform weighting schemes. We pay particular attention to the precision and accuracy of the recovered power spectra and use a Fisher information matrix formalism to examine the implications for the precision and accuracy of cosmological parameter constraints.

The remainder of the paper is organised as follows. Section \ref{sec:methodology} details the methodology, including details on the simulated weak lensing and inverse-variance weight maps and interpretation of cosmological forecasts. Section \ref{sec:results} presents the results, and Section \ref{sec:interpretation} interprets these in light of current applications of the pseudo-$C_\ell$ estimator. Finally, we present conclusions in Section \ref{sec:conclusions}.

\section{Methodology}\label{sec:methodology}

We wish to test the precision and accuracy of the pseudo-\Cell estimate for the shear-shear, position-shear and position-position spectra for various choices of pixel weighting scheme. To do so, we produce a number of realisations of correlated shear and position fields which may be analysed using different weight maps. We discuss the process by which we produce and analyse these simulations here.

\subsection{The pseudo-\Cell estimator}

In this section, we briefly describe the pseudo-\Cell estimator and its application to weighted fields. The discussion here is at a high level to motivate the propagation and modelling of pixel weights, and the interested reader is encouraged to consult \citet{Brown2005PsClCMB, AlonsoNaMaster, Alonso2024, Nicola21, Wolz24, Tessore24} for a more detailed discussion.

Consider a weighted shear field,
\begin{align}
    \tilde{\gamma}_1(\bmath{n}) \pm \rm{i}\tilde{\gamma}_2(\bmath{n}) & \equiv w(\bmath{n})\left[\gamma_1(\bmath{n}) \pm 
    {\rm i}\gamma_2(\bmath{n})) \right] \, \nonumber \\
    & = \sum_{\ell m}\left(\tilde{\mathbfss{E}}_{\ell m}\pm {\rm i}\tilde{\mathbfss{B}}_{\ell {m}} \right)\,_{\pm 2}\mathbfss{Y}_{\ell {m}}(\bmath{n})\, ,\label{eqn:spherical_gamma}
\end{align}
where quantities with tildes indicate the shear field, $\mathbf{\gamma} = \gamma_1 + \rm i \gamma_2$, multiplied by a spatially varying weight function, $w$, both evaluated at position $\hat{\mathbf{n}}$. The final line gives a spin-2 decomposition of the weighted shear field into the curl-free pseudo $E$-mode ($\tilde{E}_{\ell m}$) and gradient-free pseudo $B$-mode ($\tilde{B}_{\ell m}$) harmonic coefficients, analogous to the $E$ and $B$ components of an electromagnetic field. The $_{\pm 2}Y_{\ell {m}}$ are the spin-2 weighted spherical harmonics \citep{newman1966}. The weight function can be used to capture sky coverage effects, with pixels set to zero where observations do not exist and/or are rejected, but may formally be any non-stochastic, non-binary function which is uncorrelated with the shear. As such, as well as capturing the effects of a cut sky, it may also be used to apply an arbitrary weighting scheme in the analysis of the shear map.

Corresponding power spectra for the $E$-modes, $B$-modes, and their cross-correlation can be constructed from
\begin{align}
\tilde{\bmath{C}}^{EE}(\ell) & = \frac{1}{2\ell + 1}\sum_m \tilde{\mathbfss{E}}_{\ell m} \tilde{\mathbfss{E}}^*_{\ell m}, \\
\tilde{\bmath{C}}^{BB}(\ell) & = \frac{1}{2\ell + 1}\sum_m \tilde{\mathbfss{B}}_{\ell m} \tilde{\mathbfss{B}}^*_{\ell m}, \\
\tilde{\bmath{C}}^{EB}(\ell) & = \frac{1}{2\ell + 1}\sum_m \tilde{\mathbfss{E}}_{\ell m} \tilde{\mathbfss{B}}^*_{\ell m}.
\end{align}
Although these spectra are shown as constructed for the weighted fields, equivalent estimators can be constructed for the unweighted fields.

The weighted (pseudo) harmonic coefficients may be related to the harmonic coefficients of the true (full-sky) $E$- and $B$-fields via the spin-weighted window functions as 
\begin{align}
    \tilde{\mathbfss{E}}_{\ell m} & = \sum_{\ell' m'}\left(\mathbfss{E}_{\ell m} \tens{W}^+_{\ell\ell' m m'} + \mathbfss{B}_{\ell m} \tens{W}^-_{\ell\ell' m m'} \right ) \, , \label{eqn:pseudo_e}\\
    \tilde{\mathbfss{B}}_{\ell m} & = \sum_{\ell' m'}\left(\mathbfss{B}_{\ell m} \tens{W}^+_{\ell\ell' m m'} - \mathbfss{E}_{\ell m} \tens{W}^-_{\ell\ell' m m'} \right )\, , \label{eqn:pseudo_b}
\end{align}
where we have defined 
\begin{align}
\tens{W}^+_{\ell\ell' m m'} & = \frac{1}{2} \left(_2\tens{W}^{i}_{\ell\ell' m m'} + _{-2}\tens{W}^{i}_{\ell\ell' m m'} \right), \\
\tens{W}^-_{\ell\ell' m m'} & = \frac{1}{2} \left(_2\tens{W}^{i}_{\ell\ell' m m'} - _{-2}\tens{W}^{i}_{\ell\ell' m m'} \right).
\end{align}
The harmonic space window functions are related to the spatial weight function as (here defined for tomographic redshift bin $i$)
\begin{align}
    _s\tens{W}^{i}_{\ell\ell' m m'}  \equiv & \int {\rm d}\Omega \,\;_{s}\mathbfss{Y}_{\ell' m'} (\bmath{n}) w^{i}(\bmath{n}) \,_{s}\mathbfss{Y}^*_{\ell m} (\bmath{n}), \,\label{eqn:spherical_weights}
\end{align}
where $s = (0, \pm 2)$ labels the spin of the field. From this, one can define the tomographic mixing functions as 
\begin{align}
    \tens{W}^{MN}_{\ell\ell'}(i,j) = \frac{1}{2\ell+1}\sum_{mm'}\tens{W}^{M,i}_{\ell\ell' m m'}\left(\tens{W}^{N,j}_{\ell\ell' m m'}\right)^*\, ,
    \label{eqn:mixing_matrix_w++}
\end{align}
where $M,N = (+, -, 0)$, and $W_{\ell\ell'mm'}^{0, i}\equiv{}_{0}W^i_{\ell\ell' mm'}$.

Under this framework, the expectation of the spectra of the weighted map is related to the underlying spectra as
\begin{align}
    \left\langle\tilde{\bm{C}}_{\ell}^{ij}\right\rangle = \sum_{\ell'} \tens{M}_{\ell \ell'}(i,j) \bm{C}_{\ell'}^{ij}\, ,
    \label{eqn:coupled_cls}
\end{align}
where $\tens{M}$ defines a mixing matrix which relates angular modes between the weighted field and the underlying unweighted field. By applying this relation as presented, on can forward model the impact of a weight map or mask on the power spectra predicted for any given theoretical model, to be compared directly to the coupled \Cell measured directly on a data map \citep[e.g.][]{Loureiro2022}. Alternatively, by inverting the mixing matrix one can directly decouple the measured \Cell to compare to a full sky or unweighted theory expectation.

Including now the spin-0 position field, for the angular clustering power spectra,
\begin{equation}
\tens{M}_{\ell \ell'}(i, j) = \mathbfss{W}_{\ell\ell'}^{00}(i, j).
\end{equation}
For the vector of cosmic shear spectra in order of (${E}^i{E}^j, {E}^i{B}^j, {B}^i{B}^j$):
\begin{align}
\tens{M}_{\ell \ell'}(i, j) = 
\begin{pmatrix}
\mathbfss{W}_{\ell\ell'}^{++}(i,j)     &   \mathbfss{W}_{\ell\ell'}^{-+}(i,j)+\mathbfss{W}_{\ell\ell'}^{+-}(i,j)     &   \mathbfss{W}_{\ell\ell'}^{--}(i,j)  \\
-\mathbfss{W}_{\ell\ell'}^{+-}(i,j)    &   \mathbfss{W}_{\ell\ell'}^{++}(i,j)-\mathbfss{W}_{\ell\ell'}^{--}(i,j)     &   \mathbfss{W}_{\ell\ell'}^{-+}(i,j)  \\
\mathbfss{W}_{\ell\ell'}^{--}(i,j)     &   -\mathbfss{W}_{\ell\ell'}^{-+}(i,j)-\mathbfss{W}_{\ell\ell'}^{+-}(i,j)    &   \mathbfss{W}_{\ell\ell'}^{++}(i,j)  \\
\end{pmatrix}.
\end{align}
For the vector of galaxy-galaxy lensing spectra in order of ($\delta^i{E}^j, \delta^i{B}^j$):
\begin{equation}
\tens{M}_{\ell \ell'}(i, j) =
\begin{pmatrix}
    \mathbfss{W}_{\ell\ell'}^{0+}(i,j)  & \mathbfss{W}_{\ell\ell'}^{0-}(i,j)   \\
    -\mathbfss{W}_{\ell\ell'}^{0-}(i,j) & \mathbfss{W}_{\ell\ell'}^{0-}(i,j)
\end{pmatrix}.
\end{equation}

\subsection{Simulations}\label{sec:Simulations}
 
To produce correlated galaxy over-density and shear maps, we utilise \textsc{GLASS}\footnote{\url{https://glass.readthedocs.io/stable/}} \citep[Generator for Large Scale Structure,][]{GLASS}, conditioned on shell matter power spectra constructed from \textsc{CAMB}\footnote{\url{https://camb.readthedocs.io/en/latest/}} \citep[Code for Anisotropies in the Microwave Background,][]{Lewis2000CAMB}. These shell spectra are constructed assuming a top-hat window function of width $\Delta z = 0.05$. All cosmological parameters not used in the Fisher forecast are set to the \textsc{CAMB} default\footnote{For version 1.5.4}, and fiducial values for parameters which are forecast are detailed in Section \ref{sec:fisher}.

\begin{figure}
\centering
\includegraphics[trim = {40. 73. 35. 70.}, clip, width=1.\columnwidth]{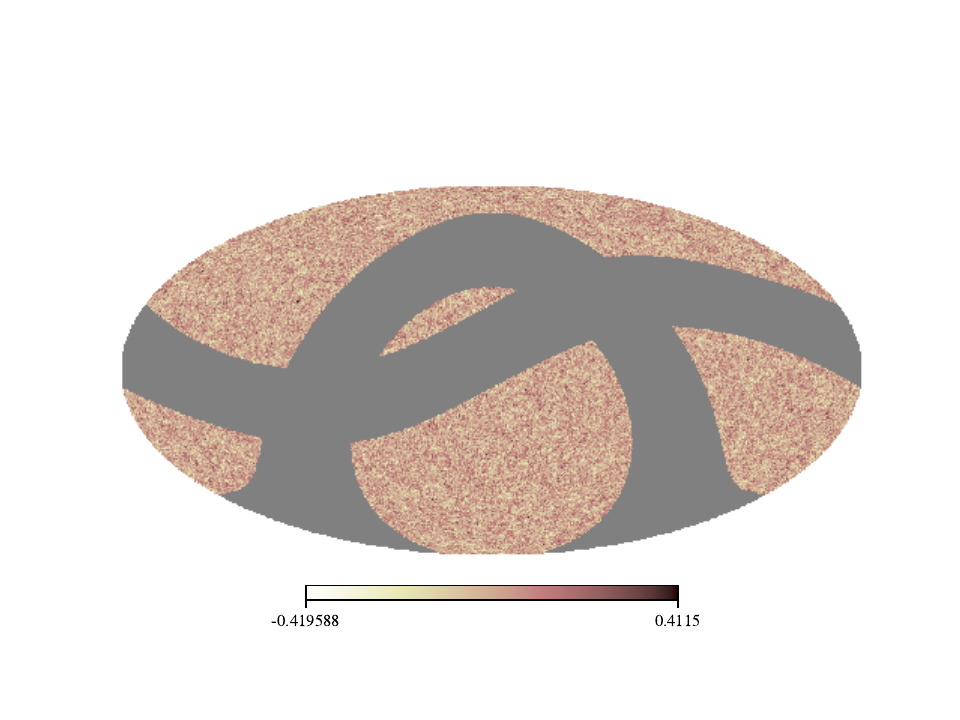}
\caption{The binary mask applied to the maps during pseudo-\Cell construction, including a basic Galactic and Ecliptic exclusion zone, superimposed on an example clustering over-density map. Gray pixels are masked, corresponding to approximately $50\%$ of the sky.}
\label{fig:mask}
\end{figure}

We model a generic current generation survey, assuming a \textsc{GLASS}-derived survey mask which contains a basic model of Ecliptic and Galactic obscuration, masking approximately 50\% of the sky (Fig. \ref{fig:mask}). We assume a galaxy redshift distribution following:
\begin{equation}
n(z) = \mathcal{N}(z_{\rm ph}, \sigma)*n(z_{\rm ph}),
\end{equation}
with photometric distribution
\begin{equation}
    n(z_{\rm ph}) \propto \left(\frac{z_{\rm ph}}{z_0}\right)^2 \exp\left(-\frac{z_{\rm ph}}{z_0}\right)^{-1.5},\label{eqn:nz}
\end{equation}
where $z_0 = z_{\rm med}/1.412$ and $z_{\rm med} = 0.7$ is the median of the survey, and $*$ denotes convolution with a photometric redshift uncertainty distribution modelled as a Gaussian with width $\sigma = 0.05(1+z_{\rm ph})$.

\begin{figure}
\centering
\includegraphics[width=\columnwidth]{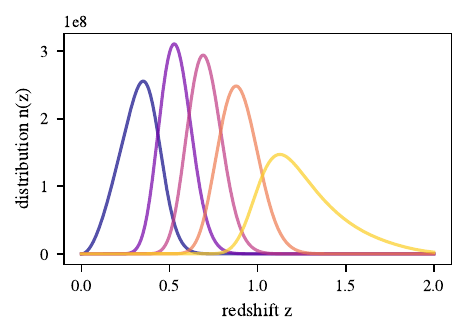}
\caption{Redshift distributions for the 5 tomographic bins considered in this analysis. Bins are chosen such that each bin contains the same number of galaxies. The ordinate axis is given in terms of the number of galaxies per steradian.}
\label{fig:nz}
\end{figure}

We construct maps according to tomographically-binned redshift distributions, covering a redshift range $0 < z < 2$, separated into 5 equi-populated bins (Figure \ref{fig:nz}), with each bin containing an average number density of galaxies of $6$ galaxies per square arcminute (giving a total of 30 galaxies per square arcminute across the full sample). Whilst this analysis could consider a larger number of redshift bins, we instead limit to 5 bins to aid interpretation and visualisation. Wong et al (in prep) shows that the information content on dark energy equation of state parameters saturates around $7$--$8$ bins.

The lensing convergence for tomographic bin $i$ is related to the underlying matter density contrast $\delta_{\rm M}$ as
\begin{equation}
\kappa^i(\bmath{n}) = \int^{\chi_H}_{0} {\rm d}\chi \;q^i(\chi) \delta_{\rm M}(\bmath{n}, \chi)
\end{equation}
where the weight is given by
\begin{equation}
q^i(\chi) = \frac{3H_0^2\Omega_{\rm M}}{2c^2}(1+z|_{\chi})\int_{\chi}^{\chi_H} {\rm d}\chi' \; n^i(z) {\rm d}z/{\rm d}\chi \frac{\chi(\chi'-\chi)}{\chi},\label{eqn:kappa_weights}
\end{equation}
where $H_0$ is the Hubble constant, $c$ the speed of light, $\chi$ the co-moving distance and $\chi_H$ the co-moving horizon, and a flat cosmology has been imposed.

The GLASS framework produces realisations of the shear field in a given redshift bin by integrating over realisations of the matter over-density field simulated in multiple redshift slices. GLASS produces a corresponding realisation of the galaxy over-density field from the same matter over-density fields used to create the shear field. In this way, realistic correlations between the galaxy over-density and shear fields will be captured, hence these simulations are suitable for testing the effect of SLC in recovered 2-point statistics. Magnification bias and obscuration effects are not explicitly included in the simulations. However, given their net effect is very similar to that of SLC (i.e. all three effects act to introduce additional correlations between the galaxy over-density and shear fields), we contend that the inclusion of the SLC effect on its own is sufficient to demonstrate the effectiveness of our suggested approach. 

To determine the statistical properties of the weighting schemes of interest, we aim to produce a large ($R=500$) suite of simulated weak lensing and clustering maps to analyse. For each new realisation, we construct a set of galaxy over-density and shear maps for each redshift bin by summing over an $n(z)$ weighted contribution from each log-normally distributed matter field shell produced with \textsc{GLASS}. The use of log-normal fields means that the galaxy over-density fields are physical across all pixels, bounded from below by $N \ge 0$ galaxy counts or galaxy over-density $\delta \ge -1$, including at low redshifts. The simulated maps are generated using the \textsc{Healpix}\footnote{\url{http://healpix.sourceforge.net/}} pixelisation scheme \citep{Healpy2}, using a \textsc{Healpix} resolution parameter, $N_{\rm side}=1024$, corresponding to a pixel scale of $\sim$3.4 arcmin. This allows for accurate power spectrum reconstruction up to multipoles $\ell_{\rm max} \approx 1500$.


The galaxy counts field has Poisson noise added as
\begin{equation}
    N(\bmath{n})\sim \mathrm{Poisson}\{\bar{N}[1+\delta(\bmath{n})]\}
\end{equation}
where $\bar{N}(1 + \delta)$ is the local galaxy counts for that pixel and given realisation, $\bar{N}$ the average number for the tomographic bin, and $\delta$ the galaxy over-density for that pixel. The shear fields have shape noise added as 
\begin{equation}
\gamma(\bmath{n}) \to \gamma(\bmath{n}) + \epsilon(\bmath{n}),
\end{equation}
with $\epsilon \sim \mathcal{N}(0, \sigma_\epsilon(\bmath{n}))$, where $\mathcal{N}$ is a Normal distribution with width
\begin{equation}
    \sigma^2_\epsilon(\bmath{n}) = \frac{\sigma^2_e}{N(\bmath{n})}. \label{eqn:shape_noise}
\end{equation}
Here, $\sigma_e=0.28$ is the shape noise, denoting the distribution of galaxy intrinsic ellipticities across the full galaxy population. The presence of  the denominator $N$ therefore indicates that in averaging over the source shears in a given pixel, the uncertainty on the pixel estimate of the mean shear is reduced according to the number of sources in the pixel.

\subsection{Map to pseudo-$C_\ell$ weighting schemes}\label{sec:weighting_schemes}

In this paper, we are concerned primarily with how one might weight pixel maps of the weak lensing shear and clustering over-density, to best recover the 3$\times$2pt power spectra, and consequently cosmological information contained therein. Consider then the case where the sky has been divided into equal-area \textsc{Healpix} pixels, each containing an occupation of galaxies from the full survey sample. An estimate of the shear within the pixels may be given by the weighted average of the shear 
\begin{equation}
    \hat{\gamma}_{\rm pix} = \frac{\sum_{i} w_i\gamma_i}{\sum_i w_i},\label{eqn:general_weighted_shear_av} 
\end{equation}
where the sum is taken over all galaxies in the sample which are present within the bounds of each pixel, and $w_i$ is a weight applied to each galaxy in that sample. The uncertainty on the pixel-averaged estimate of the shear is minimised by choosing the weighting to be the inverse of the variance of the shear estimate for each galaxy.

We can consider the variance associated with each galaxy as a combination of the uncertainty due to the distribution of intrinsic ellipticities in the global population of galaxies (termed the shape noise, $\sigma_e$), and the measurement uncertainty associated with the application of a shear measurement technique ($\sigma_{\rm meas})$ \citep[e.g.][]{Congedo2024LensMC},
\begin{equation}
    w_i = \frac{1}{\sigma^2_e + \sigma^2_{\rm i, meas}}.\label{eqn:source-shear-weight}
\end{equation}
The shape noise is constant for all sources, whilst the measurement error is source specific. In the r\'egime where the shears in a pixel are independent and identically distributed (so that the uncertainty in each is source-independent), the inverse-variance pixel weight becomes
\begin{equation}
    w(\bmath{n}) = \frac{1}{\sigma^2_{\rm pix}(\bmath{n}) }  = \frac{N(\bmath{n})}{\sigma^2_\gamma},\label{eqn:var_pixel_weight}
\end{equation}
where $\sigma^2_{\rm pix}$ is the variance of the pixel shear estimate ($\hat{\gamma}$, equation~\ref{eqn:general_weighted_shear_av}), $N$ is the number of sources in the pixel and $\sigma_\gamma$ denotes the width of the distribution of shear for the population of sources. Where the only contribution to the uncertainty on each galaxy's shear estimate is source-independent shape noise (as in Eqn \ref{eqn:shape_noise}), $\sigma_\gamma \equiv \sigma_e$. 

We see explicitly in equation (\ref{eqn:var_pixel_weight}) that inverse-variance weighting a shear map induces a sensitivity to the local number density of galaxies, and hence to the local galaxy clustering signal. We note that although this dependence is explicit in this r\'egime due to the choice to consider only the source-independent contribution to the total galaxy uncertainty (as in the case of shape noise), the dependence of the inverse-variance pixel weight to local clustering is implicit in the general case.

In the following, we consider therefore two weighting schemes:
\begin{enumerate}
    \item Inverse-variance weighting, where we consider only the contribution of the shape noise as a floor to the per-source uncertainty. In this case, Eqn (\ref{eqn:var_pixel_weight}) holds, and we impose $\sigma_\gamma = \sigma_e$. We label this scheme ``inv-var'' weighting.
    \item Unweighted, with $w_{\rm pix} = 1$ for all observed pixels. We label this scheme ``unity'' weighting.
\end{enumerate}

Full sky (decoupled) $C_\ell$ estimates for each combination of redshift bin maps are computed using \textsc{NaMaster}, for each realisation, where the weight map of Eqn (\ref{eqn:spherical_gamma}), $w(\mathbf{\hat{n}})$, is set according to the above two options (i.e. $C_\ell$'s are measured using both weighting schemes described above). We note that due to the dependence of the shear map noise and inverse-variance weight map on the local clustering signal, we cannot easily forward model the impact of the weighting scheme by coupling the theoretical $C_\ell$'s, since this will vary between realisations. Instead, we compare decoupled $C_\ell$'s to the full-sky theoretical model power spectra, appropriately binned in $\ell$-space to match the binning of the power spectra measured from the simulations. In contrast, the unity weighting can be forward modelled. However in order to present a fair comparison, we propagate the unity-weighting through the same measurement and reconstruction process as used in the inverse-variance weighted analysis. 

\subsection{Cosmological parameter uncertainty and bias forecasts}\label{sec:fisher}

As well as considering the impact of the weighting schemes on the accuracy and precision of the reconstructed power spectra, we wish to  understand their impact in terms of the accuracy and precision of cosmological parameter constraints. We therefore consider a Fisher information forecast. The Fisher information matrix is defined as \citep[e.g.][]{TegmarkTaylorFisher, Thiele2020},
\begin{equation}
\mathbfss{F}_{\alpha\beta} = -\left\langle\frac{\delta^2 \mathcal{\ln L}}{\delta \theta_{\alpha}\delta \theta_{\beta}}\right\rangle =  \bmath{M}_{,\alpha} \mathbfss{C}^{-1} \bmath{M}_{,\beta}\;, \label{eqn:Fisher}
\end{equation}
where $\bmath{M}$ is a model vector containing elements over all considered power spectra $C^{ij}_{\rm mn}(\ell)$, covering all wavenumber ($\ell$) modes, redshift bin ($ij$) and observable ($mn$) combinations. $\bmath{M}_{,\alpha}$ denotes the derivative of the model vector with respect to inferred model parameter $\theta_{\alpha}$,  $\mathbfss{C}$ is the covariance matrix for this model vector and matrix multiplication is implied. In the baseline \3x2pt experiment, the model vector covers all combinations of the observables $mn = (EE, \delta E, \delta\delta)$, where $E$ is the $E$-mode of the cosmic shear and $\delta$ is the galaxy over-density. Thus the \3x2pt model vector contains contributions from cosmic shear $E$-modes, galaxy-galaxy lensing and galaxy clustering, as standard.

As the first equality in equation~(\ref{eqn:Fisher}) implies, for a given likelihood $\mathcal{L}$, the Fisher matrix gives a good approximation of the width of a Gaussian likelihood in model space when evaluated at a fiducial parameter value which is taken to be the truth. The second equality assumes a Gaussian likelihood in data-space. We note that a Gaussian likelihood has been shown to be a good approximation to the true likelihood of power spectrum estimates and will be sufficient for precision cosmology with Stage-IV surveys \citep{upham2021, hall2022}. 

Model derivatives around the fiducial parameter set, required to evaluate equation~(\ref{eqn:Fisher}), are computed from \textsc{CAMB} using finite differences, where the step size is tuned to ensure convergence of the derivatives. We consider a default parameterisation where the cosmological parameters of interest are restricted to the dark energy equation of state and its time evolution (modelled as $w(z) = w_0 + w_a \, z / (1+z)$, with fiducial parameter values, $w_0 = -1, w_a = 0$), the cold dark matter content $\Omega_{\rm c}=0.25$ and the root-mean-square linear matter density fluctuations within an $8 \, h^{-1}{\rm Mpc}^3$ sphere ($\sigma_8\sim0.79)$. The latter parameter is a derived parameter consistent with this cosmological set-up and a primordial power spectrum amplitude of $A_s = 2\times10^{-9}$. We present results using the derived parameter $S_8 = \sigma_8\sqrt{\Omega_{\rm M}/0.3}$, where $\Omega_M=0.3$ is the total matter content of the Universe, containing both dark and baryonic contributions, and a perturbation on $S_8$ is applied by varying $\sigma_8$ whilst keeping $\Omega_{\rm M}$ fixed. We consider the impact of linear galaxy bias by marginalising over an independent galaxy bias per redshift bin, implying $5$ linear galaxy bias nuisance parameters each set for a fiducial value of $b^i = 1 \;\forall \;i \in (1,5)$. 

The covariance matrix is determined directly from the simulations themselves, for both weighting schemes, by computing the covariance across multiple realisations of the data vector (constructed with the same ordering as the model vector $M$), as recovered from the multiple \textsc{GLASS} simulations. As we measure mask-decoupled full-sky $C_\ell$ estimates from \textsc{NaMaster}, utilising identical bandpower-binning for all observables, these covariance are dominated by the block-diagonals for each observable in the data/model vector.  In the limit of a large number of realisations, noise, particularly in the block off-diagonal elements where the covariance is zero, is suppressed. However, we find that robust results are obtained for fewer realisations where the covariance is constructed such that block off-diagonals are set identically to zero, thus suppressing no- or low-signal r\'egimes where the noise is dominant.

The Fisher information matrix may also be used to forecast biases in inferred parameters as \citep[e.g.][]{Thiele2020, AmaraFisherBias} 
\begin{equation}
\Delta \theta_{\alpha} = \sum_\tau (\mathbfss{F}^{-1})_{\alpha\tau} \bmath{M}_{,\tau} \mathbfss{C}^{-1} \Delta \bmath{M},
\end{equation}
where $\Delta \bmath{M}$ represents a bias in the data vector which is not modelled in the model vector. In the following, $\Delta \bmath{M}$ is constructed as the mean over all simulated model vector realisations, minus the theoretical model vector prediction at the fiducial cosmology as calculated by \textsc{GLASS}. As such, where the recovered $C_\ell$'s are consistent with the theory prediction in their mean, this will not predict a cosmological bias in contrast to the case where the recovered $C_\ell$'s are biased.

We do not model systematic effects such as magnification or intrinsic alignments. As a result, in producing Fisher forecasts we instead consider redshift bin combinations where the background sample exists at the same or higher redshift than the foreground sample, for all observables. 

By utilising the above method, we can forecast the impact of the chosen weighting scheme on the precision and accuracy of cosmological parameters. However, we note that there are limitations to this application, beyond the usual considerations for a Fisher forecast (for example, the assumption of Gaussianity in likelihood in both data-space and model-space). We consider only linear galaxy bias, and do not model non-linear galaxy bias. We consider $C_\ell$'s up to an $\ell_{\rm max} = 1500$, irrespective of observable and redshift bin. This is likely conservative in the shear-shear correlations (which may be used to smaller scales), and optimistic for the clustering and galaxy-galaxy lensing (where more stringent cuts may be required due to the effects of the non-linear matter power spectrum and non-linear galaxy bias). We do not include intrinsic alignments in the simulations. Further, we consider a limited inferred parameter set consisting of a restricted set of cosmological parameters and linear galaxy bias and do not include nuisance parameters to account for systematic effects like shear measurement biases, redshift distribution calibration, instrinic alignments or baryonic effects.

Consequently, we emphasise that the contours presented in this paper are to be considered indicative, and not a prediction for any given survey or analysis. This is especially true when interpreting the absolute size of any parameter bias or parameter contours for a given set up. However, we also emphasise that the same assumptions are used when interpreting each weighting scheme, and thus relative parameter biases or contour areas should be considered robust subject to the survey model.

\section{Results}\label{sec:results}

\subsection{Pseudo-\Cell accuracy and precision}

\begin{figure*}
\centering
\begin{minipage}{\textwidth}
  \centering
  \includegraphics[width=.95\textwidth]{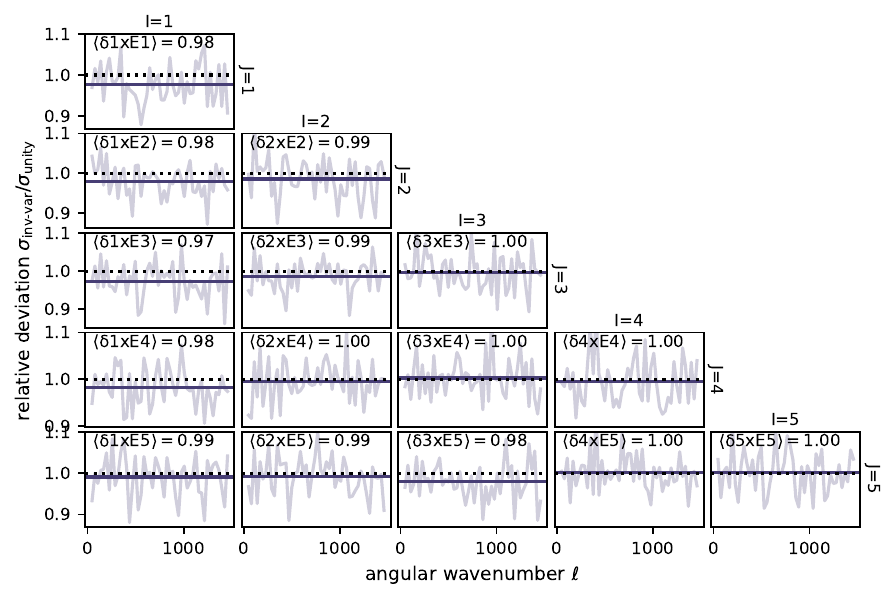}
\end{minipage}%
\\
\begin{minipage}{\textwidth}
  \centering
  \includegraphics[width=.95\textwidth]{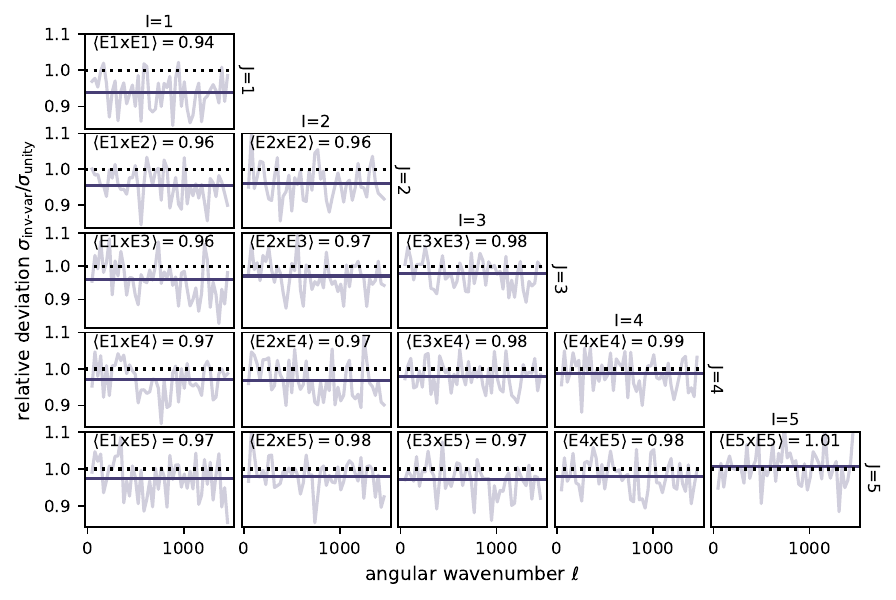}
\end{minipage}
\caption{Plots showing the ratio of statistical deviations in the estimated power spectra (defined as the root of the diagonal of the numerical covariance for each observable) between the inverse-variance and uniform weighting schemes. Values lower than 1 (dashed line) indicate that the inverse-variance weighting is returning less noisy $C_\ell$'s. Raw measurements are shown alongside the mean across all wavenumbers in dark bold solid. {\bf Top:} Position-shear. {\bf Bottom:} Shear-only.} \label{fig:cl_deviation_ratio}
\end{figure*}

In Figure~\ref{fig:cl_deviation_ratio}, we plot the ratio of the statistical deviation in the recovered power spectrum estimates using the inverse-variance weighting scheme to the statistical deviation seen in the power spectra from the uniform weighting analysis. This figure include results for all of the shear-shear ($C^{EE}_\ell$) and galaxy-galaxy lensing ($C^{\delta E}_\ell$) power spectra. (Results for the clustering spectra, $C^{\delta \delta}_\ell$, are not shown as we do not consider multiple weighting schemes for the clustering maps). In each case, the statistical deviation is calculated as the square-root of the diagonal of the block covariance matrix which describes the observable labelled. Values less than unity indicate that the inverse-variance weighting is preferred, and produces a \Cell estimate which is less noisy than that constructed without weighting shear map pixels.

We see in all cases that only a weak preference for the inverse-variance weighting scheme is seen, with at most a $~5\%$ improvement in the recovered \Cell (seen in the results for the shear-shear auto-correlation in redshift bin 1). Position-shear correlations show only a very minor improvement in the uncertainty of the recovered \Cell when using inverse-variance weighting, with the greatest improvement of $2\%$ seen in the correlation for foreground and background redshift bin 1. 

For the shear-shear correlations (lower panel), the largest improvements when using inverse-variance weighting are seen for the lowest-redshift foreground bin. For all samples cross correlating with foreground redshift bin 1, we see that the improvement is maximised for background redshift bin 1, with the improvement decreasing with increasing redshift of the background sample. This trend is consistent with expectations: the improvement expected from the use of the inverse-variance weighting scheme is maximised when the weights map has the largest range of values across the sky. Since the uncertainty on any given pixel shear value is in part driven by the local clustering (as the shear in a pixel is more precisely determined when there are more galaxies observed in the pixel, see Section \ref{sec:weighting_schemes}), then the low redshift sample would show a greater variation in the pixel weights associated with the shear map due to the greater clustering of galaxies in the late Universe. By contrast, at higher redshifts where galaxies are less clustered and with less clustering power, the distribution of inverse-variance weights is more sharply peaked reducing the effectiveness of the inverse-variance pixel weighting.

\begin{figure*}
\centering
\begin{minipage}{\textwidth}
  \centering
  \includegraphics[width=.95\textwidth]{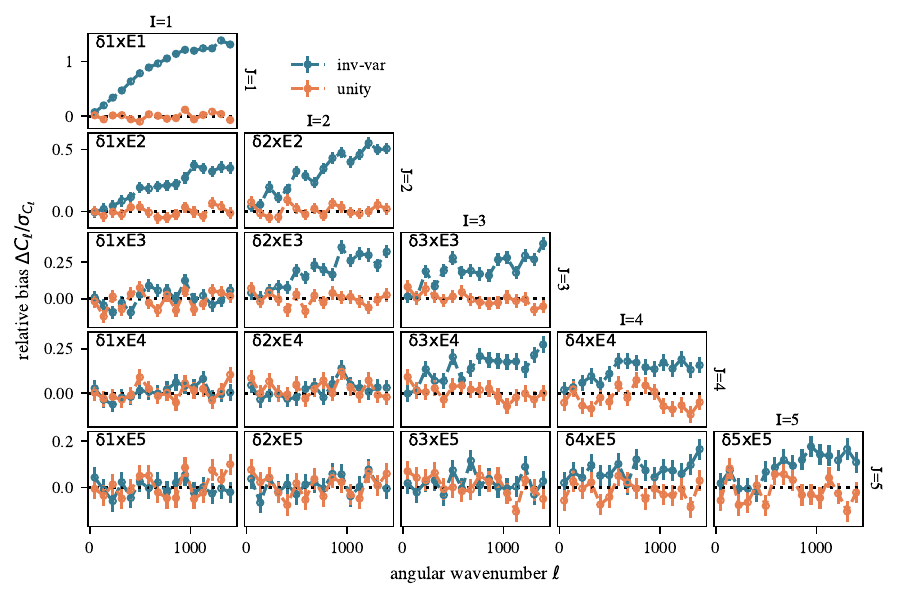}
\end{minipage}%
\\
\begin{minipage}{\textwidth}
  \centering
  \includegraphics[width=.95\textwidth]{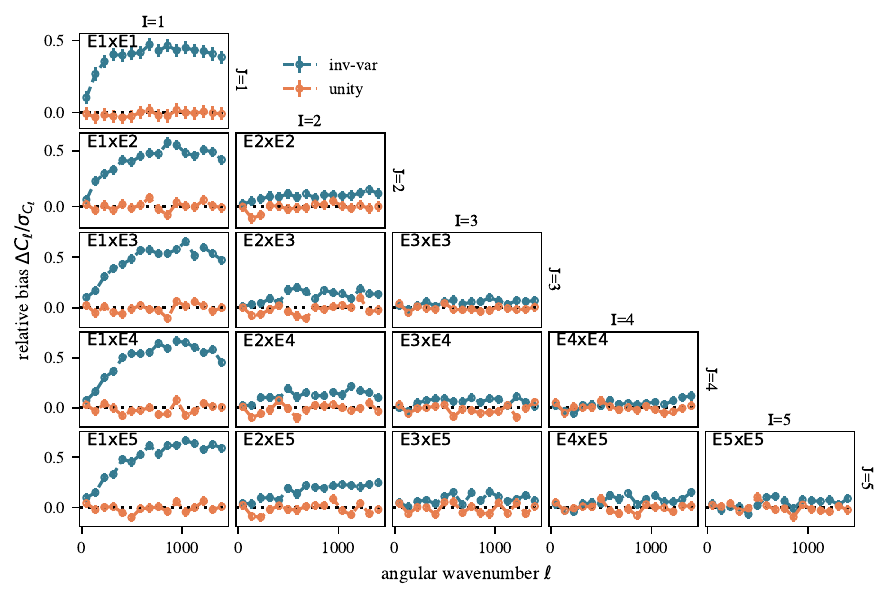}
\end{minipage}
\caption{Plots showing the significance of any bias in the recovered $C_\ell$'s (defined as the mean of the simulated $C_\ell$'s minus the theory), relative to the uncertainty in the $C_\ell$'s (defined as the root of the diagonal of the covariance for that observable.). Results from the use of inverse-variance weighting are shown in blue, and unity weighting shown in orange. For clarity of the plot, only every 3rd bandpower is shown. {\bf Top:} Position-shear. {\bf Bottom:} Shear-only.}\label{fig:cl_bias_significance}
\end{figure*}

Figure \ref{fig:cl_bias_significance} shows the bias of the recovered \Cell as a fraction of the statistical uncertainty of the recovered \Cell for each weighting scheme. The top panel shows all redshift bin combinations for correlations between a foreground position sample and a background shear sample. The bottom panel shows this for shear-shear sample correlations. The bias in recovered \Cell is determined as $\langle C_\ell \rangle - C^{\rm theory}_{\ell}$, where the first term averages over all realisations of the estimated \Cell, and the second term the expected theory \Cell.

We see that for all observables, the unweighted maps (``unity'') give unbiased estimates of the theory \Cell across all scales\footnote{Although not shown here, the position-position spectra were verified to be unbiased across all scales and for all redshift bin combinations}. By contrast, the inverse-variance weighted maps produce biased $C_\ell$'s in both the position-shear and shear-shear spectra. This bias is significant, forecast to be larger than the statistical uncertainty of the recovered spectra for the position-shear correlations between the lowest redshift bin samples, and biases $\sim\!\!50\%$ of the statistical uncertainty for shear-shear spectra.

We find that the bias is most significant in the position-shear spectra where both the foreground and background sample are close in redshift, with the bias decreasing with increasing sample redshift. For shear-shear correlations, the significance of the bias is maximised when the foreground bin is at low redshift, with little evolution of the significance of the bias with changing background sample redshift.

The origin of this bias in the recovered $C_\ell$'s when using inverse-variance weighting is due to the fact that the weights map is correlated with the shear signal through the imprint of galaxy clustering, which is the same phenomenon as the SLC effect (see the discussion in Sections \ref{sec:intro} and \ref{sec:weighting_schemes}). To confirm this assertion, we re-ran these tests on simulations where the clustering map was disassociated with the shear, by utilising an independent realisation of the clustering map when cross-correlating with the shear. We also explored an alternative where the clustering map had its pixels shuffled, therefore whitening the clustering signal. In both cases, the bias in the recovered shear-shear and position-shear spectra disappeared. Together with the lack of bias seen when using the unweighted maps, this reinforces the interpretation that it is the dependence of the pixel weights on the clustering signal that is the source of the bias seen.

\subsection{Cosmological forecasts}

\begin{figure*}
\centering
\begin{minipage}{.5\textwidth}
  \centering
  \includegraphics[width=\columnwidth]{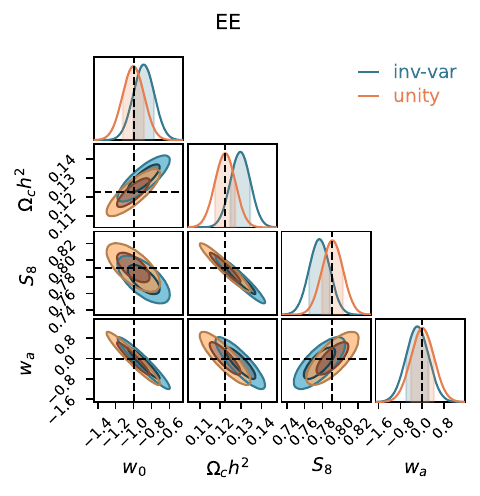}
\end{minipage}%
\begin{minipage}{.5\textwidth}
  \centering
  \includegraphics[width=\columnwidth]{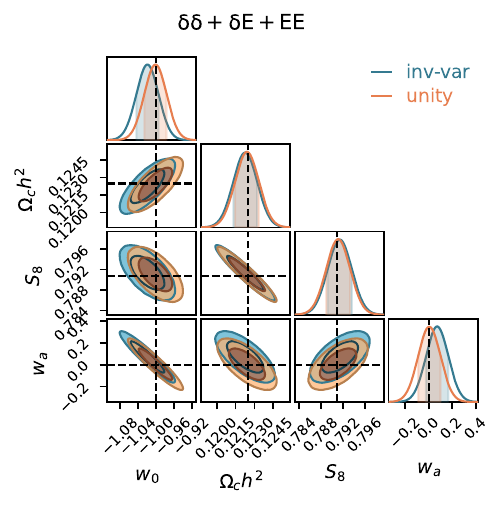}
\end{minipage}
\caption{Fisher constraints on dark energy and matter clustering cosmological parameters for the survey models described in this paper. The case where inverse-variance weighting is used in the $C_\ell$ estimation is shown in blue, and the case using a uniform weighting scheme is shown in orange. Horizontal and vertical dashed lines show the input true cosmology. {\bf Left:} Shear-only. {\bf Right:} 3$\times$2pt.}\label{fig:contours}
\end{figure*}

\begin{table*}
\centering
\caption{Table summarising the results of the Fisher forecasts. Shown are the biases on inferred parameters ($b^{p}_{w}$, where $p$ labels a cosmological parameter, and $w$ a weighting scheme), marginalised 1D parameter uncertainties ($\sigma^p_w$) and the probability-to-exceed for the maximum likelihood point in the inverse-variance weighted inference compared to the unity weighted inference. All galaxy bias nuisance parameters are marginalised over, but included in the determination of the parameter biases. As ``unity'' weighted constraints are demonstrated to be unbiased in Figure \ref{fig:contours}, we do not quote values for them here.}
\label{tab:fisher_forecasts}
\begin{tabular}{lcccccccccccccccccc}
\hline
Probe & $b^{w_0}_{\rm inv-var}$ & $b^{\Omega_ch^2}_{\rm inv-var}$ & $b^{S_8}_{\rm inv-var}$ & $b^{w_a}_{\rm inv-var}$ & $\sigma^{w_0}_{\rm inv-var}$ & $\sigma^{\Omega_ch^2}_{\rm inv-var}$ & $\sigma^{S_8}_{\rm inv-var}$ & $\sigma^{w_a}_{\rm inv-var}$ & $\sigma^{w_0}_{\rm unity}$ & $\sigma^{\Omega_ch^2}_{\rm unity}$ & $\sigma^{S_8}_{\rm unity}$ & $\sigma^{w_a}_{\rm unity}$ & p-value \\
\hline

EE & 1.1e-01 & 7.2e-03 & -1.4e-02 & -1.8e-01  & 1.2e-01 & 4.8e-03 & 1.1e-02 & 4.1e-01 & 1.2e-01 & 4.9e-03 & 1.2e-02 & 4.2e-01 & 0.005 \\
\3x2pt & -1.9e-02 & -2.3e-04 & 5.8e-04 & 7.2e-02 & 2.5e-02 & 9.5e-04 & 2.2e-03 & 9.4e-02 & 2.5e-02 & 9.4e-04 & 2.2e-03 & 9.4e-02 & 0.80 \\

\hline
\end{tabular}
\end{table*}

Figure \ref{fig:contours} shows Fisher forecasted constraints on dark energy and matter clustering cosmological parameters for the recovered $C_{\ell}$'s (as detailed in Section \ref{sec:fisher}). The left panel shows constraints for an analysis using only shear information, whilst the right shows constraints combining shear and position information in a \3x2pt analysis. In both panels, blue contours show the results from analysing $C_{\ell}$'s estimated from inverse-variance weighted maps, and orange from unweighted maps. In interpreting these figures, we remind the reader of the limitations of the forecasts detailed in Section \ref{sec:fisher}. These results are summarised in Table \ref{tab:fisher_forecasts}, including details of the probability to exceed (p-value), calculated for the biased maximum-likelihood point forecast for the inverse-variance weighted analysis in comparison to the unity-weighted contours. The p-value stated gives the probability of inferring parameters from the ``unity'' contours consistent with a shift from the maximum-likelihood larger than the forecast bias in the ``inv-var'' parameters due to statistical chance. A lower p-value therefore indicates a more statistically significant bias.

We see that the precision of inferred parameters is not impacted significantly between weighting schemes. This indicates that even though parts of the data vector show modest improvement in the precision of their recovery using inverse-variance weighting as shown in Figure \ref{fig:cl_deviation_ratio}, this does not translate into the precision of the cosmological model, as the biggest improvements are limited to low redshift bins which are the least constraining for cosmological parameters using weak lensing data. By contrast, we see that the systematic error in the recovery of the estimated $C_{\ell}$'s when using inverse-variance weighted maps may translate to a statistically significant bias in cosmological parameters, though we again caveat that the significance of this bias in any given analysis will be strongly dependent on the freedom for model variation, nuisance parameterisation and limits on the inclusion of scales (see Section \ref{sec:fisher}). In both the shear-only ($EE$) and \3x2pt cases, we see that the probability-to-exceed indicates a significant bias, especially for the shear-only which indicates a bias that exceeds statistical uncertainty for the experiment as detailed, but also for the \3x2pt which indicates a $\sim0.25\sigma$ cosmological bias.

Taken altogether, the results presented in Figures \ref{fig:cl_deviation_ratio}, \ref{fig:cl_bias_significance} and \ref{fig:contours} indicate a consistent conclusion: the use of inverse-variance weighting when constructing pseudo-$C_{\ell}$'s on shear maps gives no notable statistical advantage in $C_{\ell}$ measurement or interpretation via cosmology, but does introduce a significant bias in the recovered $C_{\ell}$'s, which could potentially lead to incorrect inference of cosmological parameter values. In the example shown, this bias is driven by the imprint of the cosmological clustering on the uncertainties of the pixelised shear estimates, and thus on the inverse-variance weight maps themselves. This leads to additional correlations in the position-shear and shear-shear power spectra, which are then not accounted for, or corrected, in the pseudo-$C_\ell$ analysis. 

\section{Interpretation and Discussion}\label{sec:interpretation}
To understand how lensing bias (be it due to SLC, magnification or obscuration effects) propagates to cosmic shear and 3$\times$2pt statistics, it is helpful to first consider the standard estimator for the 2-point shear correlation functions. To construct these, the shear for both galaxies within each galaxy pair is re-cast in terms of a tangential shear component, $\gamma_t$, and a cross-component, $\gamma_\times$ relative to the separation vector of the two galaxies. Labelling galaxies within each pair $\{\alpha, \beta\}$, the standard correlation functions estimator is 
\begin{equation}
\hat{\xi}_\pm (\theta) =  \frac{\sum_{\alpha \beta} \left[ \gamma_{t,\alpha} \gamma_{t,\beta} \pm  \gamma_{\times,\alpha} \gamma_{\times,\beta} \right] w_\alpha w_\beta \Delta_{\alpha\beta}}{\sum_{\alpha \beta} w_\alpha w_\beta \Delta_{\alpha\beta}},
\label{eqn:corrfn}
\end{equation}
where $\{w_\alpha, w_\beta\}$ are per-galaxy weights and $\Delta_{\alpha \beta} = 1$ if the separation of the two galaxies falls within the angular bin centred on $\theta$ and $\Delta_{\alpha \beta} = 0$ otherwise. If one now considers an imaginary survey containing two widely separated, equal areas -- one ``high mass'' region and one ``low mass'' region -- then, due to lensing bias one would expect the high-mass region to contain more galaxies than the low mass region.\footnote{In fact, the SLC, magnification and obscuration effects can combine in complicated ways -- e.g. in a high-density region SLC will increase galaxy numbers whereas obscuration effects will decrease them. The exact way in which all three effects combine will be complicated and highly survey-specific.} Consequently, there will be more pairs of galaxies contributing to the shear correlation sum (equation~\ref{eqn:corrfn}) from the high mass region than from the low-mass region. The resulting mean correlation function will consequently be biased high due to the fact that the shear in the high-mass region, which will be larger due to the higher mass, receives a larger weight in the correlation function sum. 

From the above discussion, it should be clear that lensing bias appearing in the shear 2-point correlation function is due to the fact that regions of the sky with high source number densities are more strongly weighted in the $\hat{\xi}_\pm$ construction than regions with lower source number densities. Considering now the pseudo-$C_\ell$ power spectrum estimator, which operates on weighted shear fields (see equation~\ref{eqn:spherical_gamma}), it should now be clear that applying a weighting scheme that depends on the local source number density will also be biased in the same way. This is the reason that the inverse-variance weighted pseudo-$C_\ell$ estimator suffers from lensing bias. The approach we advocate in this paper -- measuring two-point statistics from shear maps using uniform weights -- explicitly removes the up-weighting of regions with high galaxy number density and down-weighting of low number density regions that is embedded in the standard correlation function estimator. All that is required is that the summary estimate for the shear used in a uniform-weighted pseudo-$C_\ell$ estimate is an unbiased representation of the true underlying shear for all pixels. This requires careful determination of the pixelised shear estimates \citep{HallTessore25}.

Note that it is the projection of the galaxy shear estimates into pixelised maps and the uniform weighting of those maps in the subsequent extraction of 2-point statistics that is the important aspect of our advocated approach. Such a procedure is easy to implement for all two-point statistics and is not restricted to power spectrum analysis. We expect that it will perform equally well in terms of suppressing lensing bias effects in correlation function measurements. 

The lensing bias effects considered in this paper are not an exhaustive list: any situation where the weight function associated with a pixel is correlated with either the measured shear or clustering for that pixel can be expected to induce a bias in the recovered $C_\ell$'s. An example of this might be found in the tendency to measure galaxy shears with higher signal-to-noise when they are correlated with the point-spread-function anisotropy \citep{FenechContiKiDS, BernsteinJarvis2002, Kaiser2000}, causing a correlation between the per-source shear weights (equation \ref{eqn:source-shear-weight}) and the local shear. Such a bias would not be removed using the approach advocated here, since in this case the correlation between the per-source shear estimate and its associated weight would induce a bias in the pixelised shear field (equation \ref{eqn:general_weighted_shear_av}).

A secondary effect is where the effective shear or convergence in a given pixel may be biased due to spatial variation in the distribution of sources across a given redshift bin, either due to the clustering of sources or galaxy sample selection. In this case $n(z)\to n(z, \bmath{n})$ in equation (\ref{eqn:kappa_weights}). This additional spatial dependence can induce an additional contribution to source-lens clustering by biasing the shear maps themselves, and the recovered $C_\ell$ may be biased compared to a model that propagates the spatially-averaged $n(z)$. This additional source of SLC may be simulated by constructing catalogues, appropriately weighting the shear fields in shells using \textsc{GLASS} or in n-body simulations as in \cite{Yu2015}, where the authors demonstrate that their pixel-based estimator displays a bias due to SLC. Further, while this contribution may be expected to reduce for narrower redshift bins, as noted in \cite{Yu2015} this is limited by the photometric uncertainty in source redshift determination. The impact of this effect on the case of inverse-variance weighted pixel maps, as well as the potential mitigation through tomographic sample selection, will be investigated in a future project.

Finally, we note that this investigation has modelled a linear galaxy bias both in the production of the simulations and cosmology forecasts, with fiducial value $b = 1$ for all redshift bins. The SLC contributions to the $C_\ell$ are sourced from combinations of the source over-density, together with the lens over-density and source shear, and therefore the bias in the $C_\ell$s observed is dependent on products of the galaxy bias through the former. In \cite{DESY3Clustering}, the measured linear galaxy bias ranges between $b = (1.5, 2.3)$ across all redshift bins and measurement processes. These larger galaxy bias values would propagate to larger $C_\ell$ biases (as demonstrated also in \cite{Yu2015}), strengthening the conclusions of this investigation. We verified this by re-running a set of simulations with $b^i = 2 \;\forall\; i$ and found that whilst the uniform weighting scheme remained unbiased, the relative biases in the inverse-variance weighting scheme increases by a factor of $\sim 2$ across all wavenumbers.

\subsection{Connection to catalogue-based pseudo-\Cell estimation}

In this paper, we have presented results from pixel-level weighting schemes in the context of pseudo-\Cell estimators. Such a scheme necessarily requires that a pixelised map of galaxy shear (and over-density) is produced. However, discrete, catalogue-based pseudo-\Cell estimators have recently been proposed in \citet{Wolz24} and \citet{Tessore24}.  

By changing the pseudo-$C_\ell$ construct from a per-pixel to a per-galaxy basis, these catalogue-level estimators will unavoidably re-introduce the effect whereby high number density regions are given more weight than low-number density regions in the measured shear correlations. Thus, we expect them to be susceptible to lensing bias effects in the same way as for the standard correlation function or inverse-variance weighted pseudo-$C_\ell$ estimators. 

Further insight is gained by comparing with the map-based estimator of \cite{Tessore24}.
In that paper, the authors suggest measuring the power spectrum from weak lensing maps constructed using pixel-shear estimates, defined as,
\begin{equation}
    \hat{\gamma}_{\rm pix}^{\rm T} = \frac{\sum_{i} w_i \gamma_i}{K},
\end{equation}
where we use $\hat{\gamma}_{\rm pix}^{\rm T}$ to distinguish this from the general weighted average presented in equation (\ref{eqn:general_weighted_shear_av}), and where the sum is over all galaxies $i$ in the pixel. Importantly, the authors motivate that the denominator $K$ is a constant field mean pixel weight, which does not vary between pixels. This pixel-shear estimate can then be re-written as 
\begin{equation}
\hat{\gamma}_{\rm pix}^{\rm T} = \frac{\sum_i w_i}{K}\frac{\sum_i w_i\gamma_i}{\sum_i w_i} = \frac{1}{K \sigma^2_{\rm pix}} \hat{\gamma}_{\rm pix},
\label{eqn:tessore_equivalence}
\end{equation}
where $\hat{\gamma}_{\rm pix}$ is the general weighted average given in equation (\ref{eqn:general_weighted_shear_av}), and $\sigma_{\rm pix}^2$ in the denominator is its variance (see equation \ref{eqn:var_pixel_weight}). 

As such, we see explicitly that the map-based pseudo-\Cell estimator of \cite{Tessore24} is the same as the inverse-variance weighted estimator considered in this paper (up to a constant prefactor $K$ that is corrected in the \Cell estimate), and consequently we can interpret their discrete, catalogue-level, estimator in comparison to this.

Figure 8 of \citet{Tessore24} shows a comparison between their map-based and discrete \Cell estimators. There the authors also show a bias on small scales (large $\ell$) in their shear-shear and position-shear estimators, similar to those found here, though they find biases which are typically smaller in relation to the uncertainty in the \Cell than here\footnote{Note that in \citet{Tessore24} the survey model is different to this paper. In particular, they choose two Gaussian redshift bins at z = 0.5, 1.0 with smaller photometric uncertainty. The redshift bins 2 and 4 in this paper are the closest match, though not exact. Here, we include bins at lower redshift where the impact on pseudo-$C_\ell$ is larger. They also use a Euclid DR1-like mask covering $\sim94\%$ of the sky, compared to $\sim50\%$ here, which would increase the denominator $\sigma$, suppressing the significance of their measured bias.}. In that paper, the authors also show that by dissociating the shear and clustering maps (by taking a new independent realisation of the clustering map to combine with the shear), that bias is removed, as we also find here. As a result, the authors attribute this bias to the SLC effect, which agrees with our interpretation of the inverse-variance weighted pseudo-$C_\ell$ results presented in this paper.  

Importantly, they also demonstrate that the discrete estimator results in an almost identical bias to the map-based (inverse-variance weighted) estimate for both the position-shear and shear-shear spectra. This implies that the discrete estimator as presented is also sensitive to this same effect. 

A further observation from Fig.~8 of \citet{Tessore24} is that any residual biases associated with the use of pixels in the standard pseudo-$C_\ell$ estimator are far smaller than the SLC bias seen in the discrete estimator. Given the pixel-based, uniformly weighted pseudo-$C_\ell$ estimator effectively removes this SLC bias (and will do the same for the other lensing biases), we recommend that cosmic shear and 3$\times2$pt analyses should adopt this approach in preference to catalogue-based estimators unless the latter can be adapted to avoid lensing bias effects. 

In \citet[][]{Wolz24}, the authors do not observe a bias in the catalogue-level pseudo-$C_{\ell}$ when applied to their mock shear catalogues. However, we note that in producing the realisations of their simulated catalogues, they match each new realisation of the spin-2 shear map to galaxies sampled from the static KiDS-1000 catalogue for photometric redshift bin 4 used for galaxy positions. Whilst this process gives realistic clustering in their simulated catalogues, doing so effectively dissociates the clustering from the shear maps, which acts to artificially remove the SLC effect from the simulations. The demonstrations of the catalogue based estimator in \cite{Wolz24} are therefore closer in form to the test cases used in this paper, and in \cite{Tessore24}, to demonstrate the lack of SLC bias for dissociated shear and clustering maps.

An interesting edge case in the application of a uniform-weighted map-based estimator is where the pixel size is chosen to be sufficiently small that the typical pixel occupation is close to one. In this case, the mask itself becomes dependent on the specific realisation of the clustering field, and the data vector of pixel values is functionally similar to a catalogue of discrete shears. In this limiting case, one can expect the uniform weighted map-based estimate to re-gain sensitivity to lensing bias, due to the fact that the pixelisation has not acted to smooth out spatial variations in the number of shear estimates per pixel. This problem is avoided if the pixels are chosen to be sufficiently large such that the typical occupation of a given pixel is greater than one across the sky. As part of this analysis, we compared a subset of simulations produced with nside=2048 (compared to nside=1024 as presented here), and found that the relative bias was recovered to the same accuracy for both weighting schemes on the scales included in this analysis.

Obviously any analysis based on the use of estimators that remain susceptible to lensing bias effects require those effects to be forward modelled to avoid negatively impacting cosmological inferences. Not only does this introduce an additional model-dependence, and associated uncertainty, into the analysis but it  also significantly increases the level of complexity required in the theoretical modelling. Alternatively, the lensing bias terms may be absorbed into a sufficiently flexible nuisance parameterisation \citep[e.g.][ which demonstrated that the SLC effect may be part absorbed into intrinsic alignment and source redshift calibration parameters]{Linke2024SLC}, at the expense of complicating their interpretation and setting principled priors. This in turn may reduce cosmological parameter precision.

\section{Conclusions}\label{sec:conclusions}

In this paper, we have considered the impact of utilising an inverse-variance weighting scheme in the production of \Cell estimates from correlated galaxy clustering and weak lensing shear maps. By running a large number of shear and clustering simulations using \textsc{GLASS} \citep{GLASS}, we have assessed the impact of using an inverse-variance weighting scheme on the accuracy and precision of a pseudo-$C_\ell$ 3$\times$2pt analysis on a cut sky, in comparison to the case where the shear maps are uniformly weighted. All of our power spectrum reconstructions were performed using \textsc{NaMaster} \citep{AlonsoNaMaster}. Numerical \Cell covariance estimates were produced directly from the suite of simulated $C_\ell$'s, and \Cell biases for each scheme were produced by comparing the recovered average \Cell estimate to expected theory curves from \textsc{GLASS} and \textsc{CAMB}. 

To guide the interpretation of these results, we propagate these covariance and biases in the recovered $C_\ell$'s to a set of cosmological parameters using a Fisher information matrix formalism. This gives indicative interpretation of the results in the context of an inferred model, including both cosmological parameters and nuisance galaxy bias parameters.

In comparison to a uniform weighting, we find that the use of the inverse-variance weighting scheme gives only marginal improvement in the statistical uncertainty of the recovered \Cell across multiple redshifts and angular scales (multipoles, $\ell$). The largest improvements are found in low redshift bins, in the shear-shear signal, consistent with the imprint of clustering signal in the weights map described by clustering spectra with larger amplitudes at low redshift. The largest improvement is limited to $\sim 5\%$ in the auto-correlation of the shear sample in redshift bin 1 across all $\ell$ ranges. These marginal improvements in the precision of the recovered \Cell do not translate to improvement in the precision of cosmological parameters, with the inverse-variance weighting improving dark-energy equation of state, matter content and clustering parameter constraints to percent level.

We find, however, that use of the inverse-variance weighting scheme induces a significant bias on the recovered $C_\ell$'s, particularly when correlating lower-redshift bins across the shear and position samples.  
This bias can be a significant fraction of the statistical precision of the \Cell recovery, and is larger than this precision for the position-clustering auto-spectra, for the lowest redshift bin, on the smallest angular scales considered. This translated to significant bias in the recovered cosmological parameters in the analysis of the shear-shear spectra for the experimental setup considered ($> 1\sigma$). For a \3x2pt analysis, the bias in parameters was less significant ($\sim 0.25 \sigma$) but large enough to be a concern for future weak lensing surveys.

The bias observed in our inverse-variance weighted power spectrum analysis is due to the SLC effect \citep{Bernardeau1999SLC, Hamana2002SLC, Schmidt2009SLC, Yu2015, Deshpande2024, Linke2024SLC}. We have demonstrated that this bias can be reduced to negligible levels by instead using a uniform weighting scheme. Though we have demonstrated our approach using the pseudo-$C_\ell$ power spectrum estimator, it can be easily implemented in the context of other popular 2-point statistics, including the real-space correlation functions. 

What makes the uniform-weighting approach robust against the SLC effect is the fact that it explicitly removes, from the 2-point estimator constructions, any dependence of the weighting scheme on the local (observed) galaxy density field. We therefore expect it to be equally effective in suppressing the biases due to magnification \citep{turner1984, Schmidt2009SLC} and source obscuration \citep{hartlap2009} effects. Given that these effects, in addition to the SLC effect, are amongst the most concerning systematics for weak lensing and 3$\times$2pt analyses \citep[][]{Deshpande2024}, we anticipate this approach having significant application in the analysis of ongoing Stage-IV large-scale structure projects. 

\section*{Acknowledgements}

Some of the results in this paper have been derived using the healpy \citep{Healpy1} and HEALPix \citep{Healpy2} packages. CAJD and MLB acknowledge support from the STFC (grant number ST/X001229/1).

\section*{Data Availability}

The data underlying this article will be shared on reasonable request to the corresponding author.



\bibliographystyle{mnras}
\bibliography{optimal_pscl} 




\appendix


\bsp	
\label{lastpage}
\end{document}